\def\bq{\begin{equation}}
\def\eq{\end{equation}}
\def\bqa{\begin{eqnarray}}
\def\eqa{\end{eqnarray}}
\def\bqb{\begin{eqnarray*}}
\def\eqb{\end{eqnarray*}}
\documentstyle[12pt,epsf]{article}
\def\pr#1#2#3{ Phys. Rev. ${\bf{#1}}$ (#2) #3}

\def\pl#1#2#3{ Phys. Lett. ${\bf{#1}}$ (#2) #3}
\def\prep#1#2#3{ Phys. Rep. ${\bf{#1}}$ (#2) #3}
\def\np#1#2#3{ Nucl. Phys. ${\bf{#1}}$ (#2) #3}
\def\zp#1#2#3{ Z. Phys. ${\bf{#1}}$ (#2) #3}

\def\eg{{\it e.g.\/}}

\global\nulldelimiterspace = 0pt
 
\def\Bsl{\hbox{/\kern-.6700em$B$}} 
\def\Dsl{\hbox{/\kern-.6700em$D$}} 
\def\Wsl{\hbox{/\kern-.6700em$W$}} 

\def\roughly#1{\mathrel{\raise.3ex
    \hbox{$#1$\kern-.75em\lower1ex\hbox{$\sim$}}}}

\def\ol#1{\overline{#1}}

\def\L{ {\cal L }}
\def\O{ {\cal O }}

\def\mt{m_t}
\def\mtd{m_t^2}

\def\lam{\Lambda_{NP}}

\begin{document}
\pagenumbering{arabic}
\thispagestyle{empty}
\def\thefootnote{\fnsymbol{footnote}}
\setcounter{footnote}{1}
 
\begin{flushright}
PM/96-31 \\ THES-TP 96/10 \\
November 1996 \\
 \end{flushright}
\vspace{2cm}
%
\begin{center}
{\Large\bf Dynamical Scenarios for Anomalous Interactions
involving $t$, $b$ quarks and bosons}\footnote{Partially 
supported by the EC contract CHRX-CT94-0579.}
 \vspace{1.5cm}  \\
%
{\large G.J. Gounaris$^a$, D.T. Papadamou$^a$, F.M. Renard$^b$}
\vspace {0.5cm}  \\

$^a$Department of Theoretical Physics, University of Thessaloniki,\\
Gr-54006, Thessaloniki, Greece.\\
\vspace{0.2cm}
$^b$Physique
Math\'{e}matique et Th\'{e}orique,
UPRES-A 5032\\
Universit\'{e} Montpellier II,
 F-34095 Montpellier Cedex 5.\\

\vspace {1cm}
 
{\bf Abstract}
\end{center}
In the present paper we explore various dynamical scenarios
for New Physics (NP) which could be generated by very heavy 
particles and observed in the form of non-standard 
"low energy" interactions affecting 
the scalar sector, the gauge bosons and the quarks of the
third family. Such interactions have been previously expressed
in terms of 48 CP-conserving gauge invariant operators. We  show
that each scenario is characterized by a specific
subset of such operators. Thus, if some traces of NP are 
ever found in the
experimental data, then the comparison with the predictions of
the various scenarios, should be able to guide the search
towards the underlying dynamics  generating it.

\vspace{0.5cm}
\def\thefootnote{\arabic{footnote}}
\setcounter{footnote}{0}
\clearpage

Up to now the Standard Model (SM) has passed all tests and stands in
an amazing agreement with the experimental data \cite{SM1, SM2}.
Minor discrepancies which are occasionally announced tend to
disappear as the statistics is increasing \cite{SM2}. 
Nevertheless it is widely believed that there is some New
Physics (NP) beyond SM to be discovered, which holds the secrets
of the mysterious mechanism inducing the spontaneous breaking
of the gauge symmetry and the Higgs properties\cite{NPgen}.\par

Of course, it may turn out that the SM scalar field is just a
parameterization of something yet unknown, and that no true Higgs
particle really exists. In this context, the possibility  
that Higgs is not a particle but just a means inducing 
new strong non-perturbative interactions 
among the longitudinal $W$ and $Z$ bosons, has extensively been 
explored\cite{Chano}. This option, as well as the
exciting one that new particles associated with NP will
be produced in the present or future Colliders, are not 
addressed here. Instead, we 
concentrate on the possibility that the SM Higgs particle
really exists and will be discovered some day in a Collider, but
we assume that no other new particle will be seen in the
foreseeable future. Moreover we assume that New Physics will
appear in the form of slight modifications of the SM 
interactions among the Higgs and the other
particles appearing in the standard model \cite{NP-GLR}. \par

If the NP scale is sufficiently large, we could parameterize
these new interactions in terms of  
$dim=6$ $SU(3)\times SU(2) \times U(1)$ gauge invariant
operators \cite{Buchmuller}. 
In a recent work we have presented 
a complete list of all such operators, under the assumption that
NP is CP invariant and that it only affects the interactions
among the Higgs, the gauge bosons and the quarks of the
\underline{third} family \cite{Papadamou1}. 
In such a framework, the 
gauge boson equations of motion  cannot of course be used in order
to reduce the number of the independent NP operators, since
these equations mix-in leptons and light quarks also.
On the other hand, the
equations of motion for the scalar and the $t$ and $b$ quarks 
can still be used, since they do not mix families, provided we
neglect all fermion masses except the top.
Thus, we end up  with a list containing 
48 operators \cite{Papadamou1}. 
We have found that thirteen of these operators contribute 
at tree level to Z-peak and lower energy observables; as a result
of which eleven of them are quite strongly constrained, 
while somewhat softer constraints exist for $\ol{\O}_{DW}$,
$\ol{\O}_{DB}$, see Eqs.(\ref{listDW}, \ref{listDB}) below 
and \cite{DWDB, Hagiwara-DWDB}. 
Among the remaining operators, 32 are at most very mildly 
constrained at present, one of them called $\O_{\Phi3}$ 
gives no contribution to
any conceivable measurement, while the experimental 
constraints on the two purely gluonic operators 
($\O_G$ and $\ol{\O}_{DG}$ below) are not yet known.
In \cite{Papadamou1}, we have also given the unitarity
constraints on the aforementioned 32 operators.\par

The aim of the present paper is to investigate the 
implications and the possible 
conditions for the appearance of any such NP operator. 
We explore  various dynamical scenarios
involving naturally heavy particles which are subsequently
integrated out leaving a low energy NP interaction expressed in
terms of $dim=6$ operators. We use the same philosophy as in
\cite{Tsirigoti}, but in the present case we also include 
the possibility that gluon or quark involving operators are
generated. 
We find interesting patterns and hierarchies that are
specific to each scenario. When experimental data will be
available in the various sectors that can be tested, our results
should be useful in suggesting (selecting) what type of scenario
and quantum numbers is NP based on.\par

These dynamical scenarios are
builded following the idea that NP is caused solely by the 
scalar sector. Therefore we do not consider any
extensions of the gauge group beyond the $SU(3)\times SU(2)
\times U(1)$ level. Thus, our dynamical scenarios are just  
renormalizable models obeying   
$SU(3)\times SU(2) \times U(1)$ gauge invariance and containing,
in addition to the usual standard particles, also a "minimal"
number of new scalar and/or fermion fields whose interactions
respect (for simplicity) CP invariance. A common property
for the masses of all these fields is that they are
gauge symmetric, and may therefore be assumed to be 
sufficiently larger than the electroweak scale 
$v=(\sqrt{2}G_\mu)^{-1/2}=0.246TeV$.  \par

There are two distinct
sets of such models, depending on whether the heavy 
particles which are responsible for generating the low energy 
NP interactions, \underline{cannot} or \underline{can}
decay at tree level, to particles already existing in SM. 
These two sets lead in general to quite different NP
operators at low energies. The dominant operators induced 
from Set I come from 1-loop diagrams
involving a heavy particle running along the loop. Models of this 
kind, with either a heavy fermion or a scalar boson running 
along the loop,
have been considered in \cite{Tsirigoti} and lead to purely
bosonic NP operators. In order to generate $dim=6$ NP operators
involving quarks also, we need to enrich these models  
by assuming the simultaneous existence of at least one heavy 
boson and one heavy fermion, and give non-trivial colour to one
of these particles. 
In such a case, the dominant quark-involving NP 
operators are induced by 1-loop diagrams in which 
the heavy particle running along the loop
changes its nature from boson to fermion. \par

In the Set II of models, characterized by the fact that 
the heavy particle can
decay to standard model ones, the dominating $dim=6$
NP operators are generated
from tree diagrams involving again only heavy particle 
propagators\footnote{The possibility that some higher dimensional
operators may occasionally be dynamically enhanced will be ignored
here \cite{Arzt}.}. In this case, a single heavy boson field 
is sufficient to create $dim=6$ NP operators inducing  Higgs 
as well as  quark affecting interactions, while a 
heavy fermion can only create quark involving 
operators \cite{Arzt}. 
\par

Before starting enumerating the dynamical models, we first give
the complete list of the $dim=6$ $SU(3)_c \times SU(2)\times
U(1)$ gauge invariant operators describing in general any kind of
NP generated at high scale and affecting only the Higgs, 
the gauge bosons and the quarks of the third family. This list
contains the bosonic operators \cite{Papadamou1, Hag-boson}
\bqa
\overline{\O}_{DW} & =& 2 ~ (D_{\mu} \overrightarrow W^{\mu
\rho}) (D^{\nu} \overrightarrow W_{\nu \rho})  \ \ \
  , \ \  \label{listDW}  \\[0.1cm]  
\ol{\O}_{DB} & = & 2~(\partial_{\mu}B^{\mu \rho})(\partial^\nu B_{\nu
\rho}) \ \ \  , \ \   \label{listDB} \\[0.1cm] 
\O_{BW} & =& \frac{1}{2}~ \Phi^\dagger B_{\mu \nu}
\overrightarrow \tau \cdot \overrightarrow W^{\mu \nu} \Phi
\ \ \  , \ \  \label{listBW}  \\[0.1cm] 
\O_{\Phi 1} & =& (D_\mu \Phi^\dagger \Phi)( \Phi^\dagger
D^\mu \Phi) \ \ \  , \ \ \\[0.1cm]   \label{listPhi1}
\overline{\O}_{DG} & =& 2 ~ (D_{\mu} \overrightarrow G^{\mu
\rho}) (D^{\nu} \overrightarrow G_{\nu \rho})  \ \ \
  , \ \  \label{listDG}  \\[0.1cm]  
\O_G &= & {1\over3!}~  f_{ijk}~ G^{i\mu\nu}
  G^{j}_{\nu\lambda} G^{k\lambda}_{\ \ \ \mu} \ \ \
 ,  \ \ \   \label{listG}\\ 
\O_{\Phi 2} & = & 4 ~ \partial_\mu (\Phi^\dagger \Phi)
\partial^\mu (\Phi^\dagger \Phi ) \ \ \  , \ \ \
  \label{listPhi2} \\[0.1cm] 
\O_{\Phi 3} & = & 8~ (\Phi^\dagger \Phi) ^3\ \ \  ,
\   \label{listPhi3} \\[0.1cm] 
\O_W &= & {1\over3!}\left( \overrightarrow{W}^{\ \ \nu}_\mu\times
  \overrightarrow{W}^{\ \ \lambda}_\nu \right) \cdot
  \overrightarrow{W}^{\ \ \mu}_\lambda \ \ \
 ,  \ \  \label{listW} \\[0.1cm]  
\O_{W\Phi} & = & i\, (D_\mu \Phi)^\dagger \overrightarrow \tau
\cdot \overrightarrow W^{\mu \nu} (D_\nu \Phi) \ \ \  , \ \
\label{listWPhi} \\[0.1cm]
\O_{B\Phi} & = & i\, (D_\mu \Phi)^\dagger B^{\mu \nu} (D_\nu
\Phi)\ \ \  , \ \ \label{listBPhi} \\[0.1cm]
\O_{WW} & = &  (\Phi^\dagger \Phi )\,    
\overrightarrow W^{\mu\nu} \cdot \overrightarrow W_{\mu\nu} \ \
\  ,  \ \ \label{listWW}\\[0.1cm]
\O_{BB} & = &  (\Phi^\dagger \Phi ) B^{\mu\nu} \
B_{\mu\nu} \ \ \  , \ \ \   \label{listBB} \\[0.1cm] 
\O_{GG} & = &  (\Phi^\dagger \Phi)\, 
\overrightarrow G^{\mu\nu} \cdot \overrightarrow G_{\mu\nu} \ \
\  ,  \ \    \label{listGG} 
\eqa
where
\bq 
\Phi=\left( \begin{array}{c}
      \phi^+ \\
{1\over\sqrt2}(v+H+i\phi^0) \end{array} \right) \ \ \ \ , \ \   
\label{listPhi} 	
\eq
\bq
D_{\mu}  =  (\partial_\mu + i~ g\prime\,Y B_\mu +
i~ \frac{g}{2} \overrightarrow \tau \cdot \overrightarrow W_\mu  +
i~ \frac{g_s}{2} \overrightarrow \lambda \cdot \overrightarrow G_\mu ) \ \
 , \  \label{listDmu} 
\eq
with $v\simeq 246 GeV$, $Y$ being the hypercharge of the
field on which the
covariant derivative acts, and $\overrightarrow \tau$ and  
$\overrightarrow \lambda$ the isospin and colour
matrices applicable whenever $D_\mu$ acts on iso-doublet fermions
and quarks respectively.\par

In addition, the above list contains operators involving quarks of the
third family. These are divided into three Classes
\cite{Buchmuller, Papadamou1, topGRV1} and are given by\\ \\
{\bf Class 1.}
\bqa
\O_{qt} & = & (\bar q_L t_R)(\bar t_R q_L) \ \ \ , \ 
\label{listqt}\\[0.1cm] 
\O^{(8)}_{qt} & = & (\bar q_L \overrightarrow\lambda t_R)
(\bar t_R \overrightarrow\lambda q_L) \ \ \ ,\ \label{listqt8} \\[0.1cm]
\O_{tt} & = & {1\over2}\, (\bar t_R\gamma_{\mu} t_R)
(\bar t_R\gamma^{\mu} t_R) \ \ \ , \ \label{listtt}\\[0.1cm] 
\O_{tb} & = & (\bar t_R \gamma_{\mu} t_R)
(\bar b_R\gamma^{\mu} b_R) \ \ \ , \ \label{listtb}\\[0.1cm] 
\O^{(8)}_{tb} & = & (\bar t_R\gamma_{\mu}\overrightarrow\lambda t_R)
(\bar b_R\gamma^{\mu} \overrightarrow\lambda b_R) \ \ \ , 
\label{listtb8} \\[0.1cm]
\O_{qq} & = & (\bar t_R t_L)(\bar b_R b_L) +(\bar t_L t_R)(\bar
b_L b_R)\ \ \nonumber\\
\null & \null & - (\bar t_R b_L)(\bar b_R t_L) - (\bar b_L t_R)(\bar
t_L b_R) \ \ \ , \ \label{listqq}\\[0.1cm] 
\O^{(8)}_{qq} & = & (\bar t_R \overrightarrow\lambda t_L)
(\bar b_R\overrightarrow\lambda b_L)
+(\bar t_L \overrightarrow\lambda t_R)(\bar b_L
\overrightarrow\lambda  b_R)
\ \nonumber\\
\null & \null &
- (\bar t_R \overrightarrow\lambda b_L)
(\bar b_R \overrightarrow\lambda t_L)
- (\bar b_L \overrightarrow\lambda t_R)(\bar t_L
\overrightarrow\lambda   b_R)
\ \ \  . \label{listqq8} \\[0.1cm]
\O_{t1} & = & (\Phi^{\dagger} \Phi)(\bar q_L t_R\widetilde\Phi +\bar t_R
\widetilde \Phi^{\dagger} q_L) \ \ \ ,\ \label{listt1} \\[0.1cm]
\O_{t2} & = & i\,\left [ \Phi^{\dagger} (D_{\mu} \Phi)- (D_{\mu}
\Phi^{\dagger})  \Phi \right ]
(\bar t_R\gamma^{\mu} t_R) \ \ \ ,\label{listt2} \\[0.1cm]
\O_{t3} & = & i\,( \widetilde \Phi^{\dagger} D_{\mu} \Phi)
(\bar t_R\gamma^{\mu} b_R)-i\, (D_{\mu} \Phi^{\dagger}  \widetilde\Phi)
(\bar b_R\gamma^{\mu} t_R) \ \ \ ,\label{listt3} \\[0.1cm]
\O_{D t} &= & (\bar q_L D_{\mu} t_R)D^{\mu} \widetilde \Phi +
D^{\mu}\widetilde \Phi^{\dagger}
(\ol{D_{\mu}t_R}~ q_L) \ \ \ , \label{listDt}\\[0.1cm] 
\O_{tW\Phi} & = & (\bar q_L \sigma^{\mu\nu}\overrightarrow \tau
t_R) \widetilde \Phi \cdot
\overrightarrow W_{\mu\nu} + \widetilde \Phi^{\dagger}
(\bar t_R \sigma^{\mu\nu}
\overrightarrow \tau q_L) \cdot \overrightarrow W_{\mu\nu}\ \ \
,\label{listtWPhi}\\[0.1cm] 
\O_{tB\Phi}& = &(\bar q_L \sigma^{\mu\nu} t_R)\widetilde \Phi
B_{\mu\nu} +\widetilde \Phi^{\dagger}(\bar t_R \sigma^{\mu\nu}
 q_L) B_{\mu\nu} \ \ \ ,\label{listtBPhi}\\[0.1cm]
\O_{tG\Phi} & = & \left [ (\bar q_L \sigma^{\mu\nu} \lambda^a t_R)
\widetilde \Phi
 + \widetilde \Phi^{\dagger}(\bar t_R \sigma^{\mu\nu}
\lambda^a q_L)\right ] G_{\mu\nu}^a  \ \ \ . \ \label{listtGPhi}
\eqa\\
{\bf Class 2.}
\bqa
\O^{(1,1)}_{qq} & = &{1\over2}\, (\bar q_L\gamma_{\mu} q_L)
(\bar q_L\gamma^{\mu} q_L) \ \ \ , \label{listqq11}\\[0.1cm]
\O^{(1,3)}_{qq} & = & {1\over2}\,(\bar
q_L\gamma_{\mu}\overrightarrow\tau q_L) \cdot
(\bar q_L\gamma^{\mu} \overrightarrow\tau q_L) \ \ \ ,
\label{listqq13}\\[0.1cm]
\O_{bb} & = & {1\over2}\,(\bar b_R\gamma_{\mu} b_R)
(\bar b_R\gamma^{\mu} b_R) \ \ \ ,\\[0.1cm]\label{listbb}
\O_{qb} & = & (\bar q_L b_R)(\bar b_R q_L) \ \ \ ,
\label{listqb} \\[0.1cm]
\O^{(8)}_{qb} &= & (\bar q_L\overrightarrow\lambda b_R)\cdot
(\bar b_R\overrightarrow\lambda q_L) \ \ \ . \ \label{listqb8}
\\[0.1cm]
\O^{(1)}_{\Phi q} &= & i\,(\Phi^{\dagger}  D_{\mu} \Phi)
(\bar q_L\gamma^{\mu} q_L) -i\,(D_{\mu} \Phi^{\dagger}  \Phi)
(\bar q_L\gamma^{\mu} q_L) \ \ \ ,\label{listPhiq1}\\[0.1cm]
\O^{(3)}_{\Phi q} &=& i\,\left [( \Phi^{\dagger}
\overrightarrow \tau D_{\mu} \Phi)
-( D_{\mu} \Phi^{\dagger} \overrightarrow \tau \Phi)\right ]
\cdot (\bar q_L\gamma^{\mu}\overrightarrow\tau q_L) \ \ \ ,
\label{listPhiq3}\\[0.1cm] 
\O_{\Phi b}& = &i\,\left [( \Phi^{\dagger}  D_{\mu} \Phi)
-( D_{\mu} \Phi^{\dagger}\Phi)\right ]
(\bar b_R\gamma^{\mu} b_R) \ \ \ ,\label{listPhib}\\[0.1cm]
\O_{D b} &=& (\bar q_L D_{\mu} b_R)D^{\mu}\Phi + D^{\mu}\Phi^{\dagger}
(\ol{D_{\mu}b_R} q_L) \ \ \ ,\label{listDb}\\[0.1cm]
\O_{bW\Phi} &=& (\bar q_L \sigma^{\mu\nu}\overrightarrow \tau
b_R)\Phi \cdot
\overrightarrow W_{\mu\nu} + \Phi^{\dagger}(\bar b_R \sigma^{\mu\nu}
\overrightarrow \tau q_L)\cdot \overrightarrow W_{\mu\nu}
\ \ \ ,\label{listbWPhi}\\[0.1cm]
\O_{bB\Phi} &= & (\bar q_L \sigma^{\mu\nu} b_R)\Phi
B_{\mu\nu} + \Phi^{\dagger}(\bar b_R \sigma^{\mu\nu}
 q_L) B_{\mu\nu} \ \ \ ,\label{listbBPhi}\\[0.1cm]
\O_{bG\Phi} &=& (\bar q_L \sigma^{\mu\nu}\lambda^a b_R)\Phi
G_{\mu\nu}^a + \Phi^{\dagger}(\bar b_R \sigma^{\mu\nu}
\lambda^a q_L)G_{\mu\nu}^a \ \ \ ,\label{listbGPhi}\\[0.1cm]
\O_{b1} & = & (\Phi^{\dagger} \Phi)(\bar q_L b_R \Phi +\bar b_R
 \Phi^{\dagger} q_L) \ \ \ .\ \ \label{listb1}
\eqa\\
{\bf Class 3.}
\bqa
\O_{qB} & = & \bar q_L \gamma^\mu q_L (\partial^\nu B_{\mu\nu}) \ \
\ , \label{listqB}\\[0.1cm]
\O_{qW}& =& \frac{1}{2} \left (\bar q_L \gamma_{\mu}
\overrightarrow\tau q_L \right ) \cdot
(D_\nu \overrightarrow W^{\mu\nu}) \ \ \ , \label{listqW}\\[0.1cm]
\O_{bB} & = & \bar b_R \gamma^\mu b_R (\partial^\nu B_{\mu\nu}) \ \
\ , \label{listbB}\\[0.1cm]
\O_{tB} & = & \bar t_R \gamma^\mu t_R (\partial^\nu B_{\mu\nu}) \ \
\ , \label{listtB}\\[0.1cm]
\O_{tG}& =& \frac{1}{2} \left (\bar t_R\gamma_{\mu}
\overrightarrow\lambda t_R \right ) \cdot
(D_\nu \overrightarrow G^{\mu\nu}) \ \ \ , \label{listtG}\\[0.1cm]
\O_{bG}& =& \frac{1}{2} \left (\bar b_R\gamma_{\mu}
\overrightarrow\lambda b_R \right ) \cdot
(D_\nu \overrightarrow G^{\mu\nu}) \ \ \ , \label{listbG}\\[0.1cm]
\O_{qG}& =& \frac{1}{2} \left (\bar q_L \gamma_{\mu}
\overrightarrow\lambda q_L \right ) \cdot
(D_\nu \overrightarrow G^{\mu\nu}) \ \ \ . \ \label{listqG}
\eqa\par

We next turn to the results of the two sets of dynamical
models mentioned above. We first present the results for both
sets, and subsequently we discuss them. \\
\bigskip

\underline{Models of Set I.} The minimal models in this set
contain in addition to the standard particles, a new heavy
scalar boson $\Psi$ and a heavy left-right symmetric 
fermion\footnote{Left-right symmetry for the fermion is imposed
because it guarantees
the absence of any anomalies.} $(F_L, F_R)$ with
various colour, isospin and hypercharge specifications. Their 
quantum numbers are such that they do not allow any tree level
decay of the new heavy particles to standard model ones.
Depending on the  $\Psi$ and $F$
hypercharges, called $Y_F\equiv y$ and $Y_\Psi$ respectively, 
there are three different version-categories of the models
of Set I characterized by the Yukawa-type interactions  
$\L_j~(j=1-3)$ which couple the NP inducing heavy fields 
$\Psi$ and  $F$ to 
any of $t_R$, $q_L\equiv (t_L, b_L )$ or $b_R$; 
(see (\ref{model1}, \ref{model2}, \ref{model3}) below). 
Thus, the  interaction to
be added to the SM Lagrangian in each of these cases is
\bq
\L_{\mbox{I}}  = i\ol{F}\Dsl F-\lam \ol{F} F~+~D_\mu\Psi^\dagger
D^\mu\Psi - \lam^2 \Psi^\dagger \Psi +2g_{\Psi}(\Psi^\dagger
\Psi) (\Phi^\dagger \Phi)~+~\L_j \ \ , \label{LInew}
\eq
where we have for simplicity used a common large mass equal to
$\lam$ for both the $\Psi$ and $F$ particles. The 
assumed interaction of NP with the third family quarks 
may take the form
\bqa
\L_1 &=& f (\ol{t}_R\Psi^\dagger F_L +\mbox{ h.c.} ) \ \ \ \mbox{
for Model 1} \ \ , \label{model1}\\[0.1cm]
\L_2 &=& f (\ol{q}_L\Psi F_R +\mbox{ h.c.} ) \ \ \ \mbox{
for Model 2}  \ \ , \label{model2}\\[0.1cm]
\L_3 &=& f (\ol{b}_R\Psi^\dagger F_L +\mbox{ h.c.} ) \ \ \ \mbox{
for Model 3}  \ \ . \label{model3}
\eqa
In each of these versions, we try four different assignments for
the $\Psi$ and $(F_L,F_R)$ isospin and colour, which are
 given in Table 1.  

In this Set, the dominant $dim=6$ NP contribution arises at
1-loop level. Thus, integrating out the heavy states in 
(\ref{LInew}), we get at a
scale just below $\lam$, the NP contribution to be added
to the SM Lagrangian\footnote{For the purely bosonic operators
we have used the Seeley-DeWitt expansion explained 
in \cite{Ball}.}. This is given by
\bqa
\L_{NP} & = & \frac{1}{(4\pi \lam)^2} \Bigg \{
-c_W\, \frac{g^3}{60}\O_W ~-~c_G\, \frac{g_s^3}{30} \O_G \ 
\nonumber \\
&-& c_{DW}\, \frac{g^2}{240}\ol{\O}_{DW}
~-~ c_{DB}\, \frac{g\prime^2}{120}\ol{\O}_{DB}
~-~ c_{DG}\, \frac{g_s^2}{120}\ol{\O}_{DG} \nonumber \\
&-& c_{WW}g_\Psi\, \frac{ g^2}{4} \O_{WW}
~-~ c_{BB} g_\Psi g\prime^2 \O_{BB}
~-~ c_{GG}g_\Psi \, \frac{ g_s^2}{6} \O_{GG} \nonumber \\
&+& c_{\Phi3} g_\Psi^3 \O_{\Phi3}
~+~ c_{\Phi2}\, \frac{g_\Psi^2}{2} \O_{\Phi2}\nonumber \\
&+&  c_{tt}\, \frac{f^4}{6}\O_{tt}
~+~ c_{qq}^{(1,1)} \, \frac{f^4}{6}\O_{qq}^{(1,1)}
~+~ c_{qq}^{(1,3)} \, \frac{f^4}{6} \O_{qq}^{(1,3)}
~+~ c_{bb}\, \frac{f^4}{12} \O_{bb}\nonumber \\
&+& c_{tB\Phi}f^2\, \frac{g\prime \mt}{6\sqrt{2} v} \O_{tB\Phi}
~+~ c_{tW\Phi}f^2\, \frac{g \mt}{24 \sqrt{2} v} \O_{tW\Phi}
~+~ c_{tG\Phi}f^2\, \frac{g_s \mt}{24 \sqrt{2} v} \O_{tG\Phi}
\nonumber \\
&+& c_{t1} f^2\, \frac{\mt}{3\sqrt{2} v}\, \left [4g_\Psi + 
\left (\frac{\mt}{v} \right )^2 \right ] \O_{t1}
~-~ c_{t2} f^2\, \frac{\mtd}{12 v^2} \O_{t2}
~+~c_{\Phi q}^{(1)} f^2 \, \frac{\mtd}{6 v^2} \O_{\Phi q}^{(1)}
\nonumber \\
&-& c_{tB} f^2 \, \frac{g\prime}{6}\O_{tB}
~-~ c_{tG} f^2 \, \frac{g_s}{12}\O_{tG}
~-~ c_{qB} f^2 \, \frac{g\prime}{12}\O_{qB}
~-~ c_{qW} f^2 \, \frac{g}{24}\O_{qW} \nonumber \\
&-& c_{qG} f^2 \, \frac{g_s}{24}\O_{qG}
~-~ c_{bB} f^2 \, \frac{g\prime}{12}\O_{bB}
~-~ c_{bG} f^2 \, \frac{g_s}{24}\O_{bG} \Bigg \} \ \ , \ \label{LNP}
\eqa
where the Model dependent $c_a$-constants for the models of type
1,2,3 (see Table 1), are given in Tables 2,3,4 respectively. 
At this point it
may  be amusing to remark that the low energy NP interactions in
(\ref{LNP}) for the case of the C models with $Y_F=0$,
could arise \eg\@ if $F$ were a heavy Majorana
neutrino. We should also observe that as far as the purely
bosonic operators are concerned, the results in (\ref{LNP}) are
similar to those in \cite{Tsirigoti}.
 \par  

\noindent
\underline{Models of Set II.} As before, we only consider models
involving  scalar or a left-right symmetric fermion fields.
As in Set I, their mass terms are gauge symmetric, so that 
it is natural to assume that these particles are heavy. 
In the models of this set, the
quantum numbers of the new heavy particles are such that they
allow their decay to standard model ones at tree 
level. Because
of this, the dominant $dim=6$ NP operators are generated at tree
level and involve purely bosonic as well as quark
containing operators. The corresponding 1-loop contributions 
are ignored in Set II, since they are suppressed by a
factor of $1/(4\pi)^2$ compared to the tree-ones. 
We consider four such models,
each of which has either a new heavy scalar field 
$\Psi$ or a left-right symmetric colour-triplet 
fermion $(F_L,~ F_R)$ characterized by the isospin
$I$ and hypercharge $Y$ indicated in Table 5.

As we see from Table 5, $\Psi$ in Model 4 has the same quantum 
numbers as the standard Higgs field. For simplicity, we assume 
though that it does not mix with the usual Higgs, and that it
neither acquires a vacuum 
expectation value when gauge symmetry is broken. The interaction
which should be added to the SM Lagrangian for the models 4-6B
is given by 
\bqa
\L_4 & = & D_\mu\Psi^\dagger D^\mu\Psi - \lam^2 \Psi^\dagger \Psi 
  +f_3 [(\Psi^\dagger \Phi) (\Phi^\dagger \Phi)~+~\mbox{ h.c. }]
\nonumber \\
&+& f_1 (\ol{t}_R\widetilde \Psi^\dagger q_L + \mbox {h.c. } )
+ f_2 (\ol{b}_R \Psi^\dagger q_L + \mbox {h.c. } ) 
 + \mbox{ ... } \ \ , \label{model4}\\
\L_5 & = &i\ol{F}\Dsl F-\lam \ol{F} F~+~f_b
(\ol{b}_R\Phi^\dagger F_L + \mbox{ h.c. } ) ~+~  
f_t (\ol{t}_R\widetilde \Phi^\dagger F_L + \mbox{ h.c. } ) \ , 
\label{model5}\\
\L_{6A} & = &i\ol{F}\Dsl F-\lam \ol{F} F~+~
f_q (\ol{q}_L\Phi F_R + \mbox{ h.c. } )  \ , 
\label{model6A}\\
\L_{6B} & = &i\ol{F}\Dsl F-\lam \ol{F} F~+~
\tilde f_q (\ol{q}_L\widetilde \Phi F_R + \mbox{ h.c. }
) \ , \label{model6B} 
\eqa   
where the dots in (\ref{model4}) stand for terms
which are irrelevant for the tree diagrams dominating 
the low energy NP effective Lagrangian. Thus, 
at a scale just below $\lam$, the $dim=6$ contributions to the 
effective NP interactions are respectively given by
\bqa
\L_{NP/4} & =& \frac{1}{\lam^2}\, (f_1^2 \O_{qt}+f_2^2 \O_{qb}
+f_1f_2 \O_{qq}+f_1f_3 \O_{t1}+f_2f_3\O_{b1} +
\frac{f_3^2}{12}\, \O_{\Phi 3}) \ , \label{LNP4} \\
\L_{NP/5} &=& \frac{1}{2\lam^2}\,  \left [f_b^2 \O_{\Phi b} +
f_t^2 \left (-\O_{t2} +\frac{\mt \sqrt{2}}{v} \O_{t1} \right )
+ 2 f_tf_b \O_{t3} \right ] \ , \label{LNP5} \\
\L_{NP/6A} & =& -~\frac{f_q^2}{4 \lam^2} (\O_{\Phi q}^{(1)}
+ \O_{\Phi q}^{(3)}) \ , \label{LNP6A}\\
\L_{NP/6B} & =& \frac{\tilde f_q^2}{4 \lam^2} \left (\O_{\Phi q}^{(1)}
+ \O_{\Phi q}^{(3)} + \frac{\mt 2^{3/2}}{v}\, \O_{t1} \right) \ .
\label{LNP6B}
\eqa\par

\underline{Discussion.} Above we have considered 16 different
renormalizable models based on $SU(3)\times SU(2)\times U(1)$
gauge invariance and divided into two Sets. The models are
"minimal" 
in the sense that we are not extending the gauge group and
they are containing the minimal number of scalar and/or
fermion fields necessary in order to create some  of the
total number of the 14 bosonic and the 34 quark-involving 
$dim=6$ CP conserving gauge invariant
interactions.\par 

Concerning these models, we first observe that 14 out the total 
number of 
the 48 aforementioned operators, were never created in
any of them. These are the purely bosonic operators 
$(\O_{BW}$, $\O_{\Phi 1}$, $\O_{W\Phi}$, $\O_{B\Phi})$;
the four-quark operators $(\O_{qt}^{(8)}$, $\O_{tb}$,
$\O_{tb}^{(8)}$, $\O_{qq}^{(8)})$ from Class 1 and 
$\O_{qb}^{(8)}$ from Class 2; and the two-quark operators
$\O_{Dt}$ and $(\O_{Db}$, $\O_{bW\Phi}$, $\O_{bB\Phi}$, 
$\O_{bG\Phi})$ from Class 1 and 2 respectively. 
With respect to this, there are three remarks to be made: 
\begin{itemize}
\item The operators 
$\O_{BW}$ and $\O_{\Phi 1}$ 
 would have been created in all models of
Set I which involve an iso-doublet scalar field $\Psi$, provided
we had added to (\ref{LInew}) the renormalizable interaction
term of the form \cite{Tsirigoti}
\bq
   (\Psi^\dagger \overrightarrow \tau \Psi) \cdot
   (\Phi^\dagger\overrightarrow \tau \Phi) \ \ .
\eq
The reason we avoided  adding this term is because both,
$\O_{BW}$ and $\O_{\Phi 1}$ are quite strongly constrained by
their tree level contribution to Z-peak observables 
\cite{DeRujula, Hag-boson}. 
\item The four-quark operators $\O_{tb}$, $\O_{tb}^{(8)}$, 
$\O_{qt}^{(8)}$ and $\O_{qb}^{(8)}$,  would also be generated
if we had considered models more complicated then those
appearing in Set I. For example $\O_{tb}$ would  
be realized if Models 1A or 1B were combined with 3A or 3B; by
introducing in addition to the heavy fermion of hypercharge
$Y_F$, two iso-douplet scalars carrying hypercharges $Y_F-2/3$ and
$Y_F+1/3$ respectively. 
\item  Contrary to the above, the four-quark
operator  $\O_{qq}^{(8)}$, as well as the operators 
$\O_{W\Phi}$, $\O_{B\Phi}$, 
$\O_{Dt}$, $\O_{Db}$, $\O_{bW\Phi}$, $\O_{bB\Phi}$ and
$\O_{bG\Phi}$ are never generated in any model of the 
type appearing in both sets I and II, even if we had increased  
the number of the iso-scalar and iso-doublet fermion or
scalar fields, so far we neglect the quark masses except the top
one.
\end{itemize}\par

We next comment on the comparison between
the two Sets I and II. As we have already said, Set I  
contains models involving heavy scalars or fermions which 
\underline{cannot} decay 
to standard model particles. On the opposite, 
to Set II belong models
involving heavy scalars or fermions 
which \underline{can} decay to particles appearing
in SM. The dominant $dim=6$ NP operators are induced at tree
level in Set II and at 1-loop level in Set I. 
The two Sets produce very different
operators. Thus, the  study of the low energy structure of NP can
give  information on the responsible high energy
dynamics. In detail the relative characteristics 
of the two Sets are:
\begin{itemize}
\item Bosonic operators: The only bosonic operator
arising in Set II, is the presently un-observable 
$\O_{\Phi3}$, which depends only on Higgs self 
interactions. Of course, this operator is generated at 
tree level in Set II only when the heavy particle is a boson;
see Model 4. If the vacuum expectation value of a
heavy scalar boson involved in the first of these 
models (Model 4), 
were allowed to acquire non vanishing vacuum
expectation value, then the operators $\O_{\Phi 1}$ and  
$\O_{\Phi 2}$ would also be generated, \cite{Arzt}. 
But in any case, no anomalous triple gauge boson couplings 
can appear in
the models of Set II; compare (\ref{LNP4}-\ref{LNP6B}). 
This property would have
remained true even if we had extended the gauge group and
included new heavy gauge bosons 
coupling to SM particles at tree level.  
On the contrary, as seen from (\ref{LNP}) and the Tables 2-4, 
almost all purely bosonic operators 
may be produced in models of the type presented in Set I.
The only bosonic operators which we were
not able to produce in models of Set I are
$\O_{W\Phi}$, $\O_{B\Phi}$ \cite{Tsirigoti}.
It is also worth remarking that the models of Set II
would be very little constrained by present data in the case
that we  put $f_2 = f_b = f_q = \tilde f_q =0$ in
(\ref{model4}, \ref{model5}, \ref{model6A}, \ref{model6B}),
which would be true if NP is only induced by 
top and Higgs new interactions.   

\item 4-quark operators: The only four quark operators 
generated
in the models of Set I, are $\O_{tt}$ arising from
Models  1A-1D; $\O_{bb}$ arising from 3A-3D; and 
$\O_{qq}^{(1,1)}$ and $\O_{qq}^{(1,3)}$ produced in 
Models 2A-2D. Correspondingly, the only
four-quark operators in Set II, are $\O_{qt}$, $\O_{qb}$ and
$\O_{qq}$, all of which appear only in Model 4 characterized by
the existence of heavy scalars.
Thus, there is no overlap for the four-quark
operators produced in the two Sets. 
\item 2-quark operators of Class 1 and 2: 
The operators $\O_{tB\Phi}$, 
$\O_{tW\Phi}$ and $\O_{tG\Phi}$ appear only in Set I; 
while $\O_{b1}$, $\O_{t3}$, $\O_{\Phi q}^{(3)}$
 and $\O_{\Phi b}$  are only met
in Set II. On the other hand, the operators 
$\O_{t1}$, $\O_{t2}$ and 
$\O_{\Phi q}^{(1)}$ appear in both sets.\par
\item The operators of Class 3, (which formally are also 
two-quark operators), are only 
generated in models of Set I, and never in the 
Set II (at tree level).\par
\end{itemize}\par

Finally we comment on the possible magnitude of the NP couplings of
the various operators. As seen from (\ref{LNP}) together with 
Tables 2-4, the couplings of all the purely gauge depending 
NP operators 
$\ol{\O}_{DW}$, $\ol{\O}_{DB}$, $\ol{\O}_{DG}$, $\O_W$ and
$\O_G$ are determined by the gauge principle.
Thus, unless there is some strong non-perturbative
enhancement, it seems that there is not much freedom to make the
strength of
these operators observable. The situation looks particularly 
severe for $\ol{\O}_{DW}$, $\ol{\O}_{DB}$ and $\O_W$; while
it may be better for  $\O_G$ and $\ol{\O}_{DG}$, where the
strength is determined by the QCD coupling $g_s$, but the
background is of course larger. \par

The situation may be more favorable for the 
Higgs and quark involving operators whose strength is always 
determined by the Yukawa couplings of the
underlying dynamical theory. Since there is no a priori
constraint on these Yukawa couplings, we could hope that some of
them may be large. For example, they could become  
$f^2 \sim 4 \pi$, which is  allowed by the unitarity constraints
\cite{Papadamou1}. Note in this respect that the SM top quark
Yukawa coupling is of order one. \par

In conclusion our study has taught us several lessons:
\begin{itemize}
\item Each scenario leads to a specific selection among the list of
admissible operators. This selection is a consequence of the
quantum numbers assigned to the NP degrees of freedom. For
example, the  gluonic operators $\O_G$ and $\ol{\O}_{DG}$
appear naturally as soon as the heavy
fermion or scalar integrated out, carries colour; in complete
analogy to $\O_W$ and $\ol{\O}_{DW}$ appearing whenever the
heavy new particles carry isospin. Similarly, the  
appearance of $\O_{GG}$ and/or $\O_{WW}$ follows whenever 
the heavy particle is a scalar carrying a non-vanishing 
colour and/or isospin quantum number respectively. 
We thus emphasize that the gluonic operators
are at the same footing as the other bosonic operators,
which is a feature not considered originally
\cite{Hag-boson}. 
\item In general, a heavy fermion loop in the models of Set I, 
will only
generate purely gauge boson  operators, while  
Higgs dependent bosonic operators will appear only when 
a heavy scalar particle runs  around the loop \cite{Tsirigoti}. 
On the other hand, the generation of quark operators in
Set I needs that the heavy particle in the loop
changes its nature from  scalar to  fermion.
Correspondingly for Set II, where only tree level contributions
are considered, a diagram involving a heavy fermionic propagator
can obviously never generate  bosonic operators.
\item A natural hierarchy in the size of the couplings associated to
the involved operators is also generated. Thus we find that
there exist a set of operators which are never generated, while
other sets could appear, but with reduced strengths determined by
gauge couplings and loop-factors. On the other hand, the 
Higgs and heavy quark involving operators which are generated,
can have a strong coupling being constrained 
at present only by unitarity considerations.

\end{itemize} \par

Summarizing, we conclude that the comparison of such a
theoretical landscape with the present and future experimental
data, should be very instructive when looking for hints about
the origin and the basic structure of NP.

\noindent
\begin{center}
\begin{tabular}{|c|c|c|c|c|c|} \hline
\multicolumn{6}{|c|}{Table 1: Models of Set I.} 
\\ \hline
\multicolumn{1}{|c|}{Model} &
\multicolumn{1}{|c|}{$Y_\Psi$} &
 \multicolumn{1}{|c|}{ $I_F$} &
  \multicolumn{1}{|c|}{$I_\Psi$} &
    \multicolumn{1}{|c|}{colour$(F)$} &
     \multicolumn{1}{|c|}{colour$(\Psi)$} \\ \hline
  $1A$ & $y- 2/3$ &$1/2$& $1/2$   & $1$ & $\ol{3}$   \\ 
  $1B$ & $y-2/3 $ & $1/2$& $1/2$  & $3$  & $1$  \\ 
  $1C$ & $y-2/3 $ & $0$& $0$      & $1$  & $\ol{3}$  \\ 
  $1D$ & $y-2/3 $ & $0$& $0$      & $3$  & $1$  \\ 
  $2A$ & $1/6-y$ &$1/2$& $0$      & $1$ & $ 3$   \\ 
  $2B$ & $1/6-y$ & $1/2$& $0$     & $3$  & $1$  \\ 
  $2C$ & $1/6-y$ & $0$& $1/2$     & $1$  & $ 3$  \\ 
  $2D$ & $1/6-y $ & $0$ & $1/2$   & $3$  & $1$  \\ 
  $3A$ & $y+ 1/3$ &$1/2$& $1/2$   & $1$ & $\ol{3}$   \\ 
  $3B$ & $y +1/3 $ & $1/2$& $1/2$ & $3$  & $1$  \\ 
  $3C$ & $y+ 1/3 $ & $0$& $0$     & $1$  & $\ol{3}$  \\ 
  $3D$ & $y +1/3 $ & $0$& $0$     & $3$  & $1$  \\ \hline
\end{tabular}
\end{center}
\noindent
%
%
%
%
\begin{center}
\begin{tabular}{|c|c|c|c|c|} \hline
\multicolumn{5}{|c|}{Table 2: Non vanishing 
      $c_a$-Parameters for Models 1A-1D; (see (\ref{LNP})).} 
\\ \hline
\multicolumn{1}{|c|}{$c_a$} &
 \multicolumn{1}{|c|}{1A} &
  \multicolumn{1}{|c|}{1B} &
    \multicolumn{1}{|c|}{1C} &
     \multicolumn{1}{|c|}{1D} \\ \hline
  $c_W$  &$1$& $-5$ & $0$ & $0$   \\ 
  $c_G$ & $1$& $-2$ & $1/2$  & $-1$  \\ 
  $c_{DW}$ & $11$& $25$ & $0$ & $0$  \\ 
  $c_{DB}$ & $6\left (y-\frac{2}{3}\right )^2 +16 y^2 $
        & $ 2\left (y-\frac{2}{3}\right )^2 +48 y^2 $ 
        & $3\left (y-\frac{2}{3}\right )^2 +8 y^2 $  
        & $\left (y-\frac{2}{3}\right )^2 +24 y^2 $  \\ 
  $c_{DG}$ & $1$ & $8$ & $1/2$ & $ 4$ \\
  $c_{WW}$ & $1$ & $1/3$ & $0$ & $0$ \\ 
  $c_{BB}$ &  $\left (y-\frac{2}{3} \right )^2$
          &  $\frac{1}{3}\left (y-\frac{2}{3}\right )^2$
          &  $\frac{1}{2}\left (y-\frac{2}{3}\right )^2$ 
          &  $\frac{1}{6}\left (y-\frac{2}{3}\right )^2$ \\ 
    $c_{GG}$ & $1$ & $0$ & $1/2$ & $0$ \\
    $c_{\Phi2}$ & $1$ & $1/3$ & $1/2$ & $1/6$ \\
    $c_{\Phi3}$ & $1$ & $1/3$ & $1/2$ & $1/6$ \\
    $c_{tt}$ & $1$ & $1$ & $1/2$ & $1/2$ \\
    $c_{tB\Phi}$ & $\frac{1}{3} -y$ & $\frac{1}{3} -y$ 
                 & $\frac{1}{2}(\frac{1}{3} -y)$ 
                 & $\frac{1}{2}(\frac{1}{3} -y)$ \\
    $c_{tG\Phi}$ & $1$ & $-1$ & $1/2$ & $-1/2$ \\
    $c_{t1}$ & $1$ & $1$ & $1/2$ & $1/2$ \\
    $c_{\Phi q}^{(1)}$ & $1$ & $1$ & $1/2$ & $1/2$ \\
    $c_{tB}$ & $y+\frac{1}{3}$ & $y+\frac{1}{3}$ &
               $\frac{1}{2}(y+\frac{1}{3})$ & 
               $\frac{1}{2}(y+\frac{1}{3})$ \\
    $c_{tG}$ & $1$ & $3$ & $1/2$ & $3/2$ \\[0.2cm] \hline
\end{tabular}
\end{center}
\noindent
%
%
\begin{center}
\begin{tabular}{|c|c|c|c|c|} \hline
\multicolumn{5}{|c|}{Table 3: Non vanishing 
      $c_a$-Parameters for Models 2A-2D; (see (\ref{LNP})).} 
\\ \hline
\multicolumn{1}{|c|}{$c_a$} &
 \multicolumn{1}{|c|}{2A} &
  \multicolumn{1}{|c|}{2B} &
    \multicolumn{1}{|c|}{2C} &
     \multicolumn{1}{|c|}{2D} \\ \hline
  $c_W$  &$-2$& $-6$ & $3$ & $1$   \\ 
  $c_G$ & $1/2$ & $-2$ & $1$  & $-1$  \\ 
  $c_{DW}$ & $8$& $24$ & $3$ & $1$  \\ 
  $c_{DB}$ & $3 (\frac{1}{6}-y  )^2 +16 y^2 $
        & $  (\frac{1}{6} -y  )^2 +48 y^2 $ 
        & $ 6  (\frac{1}{6} -y  )^2 +8 y^2 $  
        & $2  (\frac{1}{6}-y  )^2 +24 y^2 $  \\ 
  $c_{DG}$ & $1/2$ & $8$ & $1$ & $ 4$ \\
  $c_{WW}$ & $0$ & $0$ & $1$ & $1/3$ \\ 
  $c_{BB}$ &  $ \frac{1}{2}(\frac{1}{6}-y)^2$
           &  $ \frac{1}{6}(\frac{1}{6}-y)^2$
           &  $ (\frac{1}{6}-y)^2$
           &  $ \frac{1}{3}(\frac{1}{6}-y)^2$ \\
    $c_{GG}$ & $1/2$ & $0$ & $1$ & $0$ \\
    $c_{\Phi2}$ & $1/2$ & $1/6$ & $1$ & $1/3$ \\
    $c_{\Phi3}$ & $1/2$ & $1/6$ & $1$ & $1/3$ \\
    $c_{qq}^{(1,1)}$ & $1/4$ & $1/2$ & $1/2$ & $1/4$ \\[0.1cm]
    $c_{qq}^{(1,3)}$ & $1/4$ & $0$ & $0$ & $1/4$ \\
    $c_{tB\Phi}$ & $\frac{1}{2}( \frac{1}{12} -y)$ 
                 & $\frac{1}{2}( \frac{1}{12} -y)$
                 & $\frac{1}{2}( \frac{1}{12} -y)$
                 & $\frac{1}{2}( \frac{1}{12} -y)$  \\
    $c_{tW\Phi}$ & $-1/2$ & $-1/2$ & $1/2$ & $1/2$ \\
    $c_{tG\Phi}$ & $1/2$ & $-1/2$ & $1/2$ & $-1/2$ \\
    $c_{t1}$ & $1/2$ & $1/2$ & $1/2$ & $1/2$ \\
    $c_{t2}$ & $1$ & $1$ & $1$ & $1$ \\
    $c_{qB}$ & $y+\frac{1}{12}$  & $y+\frac{1}{12}$ 
             & $y+\frac{1}{12}$ & $y+\frac{1}{12}$ \\ 
    $c_{qW}$ & $3$ & $3$ & $1$ & $1$ \\
    $c_{qG}$ & $1$ & $3$ & $1$ & $3$ \\[0.2cm] \hline
\end{tabular}
\end{center}
\noindent
%
%
\begin{center}
\begin{tabular}{|c|c|c|c|c|} \hline
\multicolumn{5}{|c|}{Table 4: Non vanishing 
      $c_a$-Parameters for Models 3A-3D; (see (\ref{LNP})).} 
\\ \hline
\multicolumn{1}{|c|}{$c_a$} &
 \multicolumn{1}{|c|}{3A} &
  \multicolumn{1}{|c|}{3B} &
    \multicolumn{1}{|c|}{3C} &
     \multicolumn{1}{|c|}{3D} \\ \hline
  $c_W$  &$1$& $-5$ & $0$ & $0$   \\ 
  $c_G$ & $1$& $-2$ & $1/2$  & $-1$  \\ 
  $c_{DW}$ & $11$& $25$ & $0$ & $0$  \\ 
  $c_{DB}$ & $ 6 (y+\frac{1}{3} )^2 +16 y^2 $
        & $ 2 (y+\frac{1}{3})^2 +48 y^2 $ 
        & $ 3 (y+\frac{1}{3})^2 +8 y^2 $  
        & $ (y+\frac{1}{3})^2 +24 y^2 $  \\ 
  $c_{DG}$ & $1$ & $8$ & $1/2$ & $ 4$ \\
  $c_{WW}$ & $1$ & $1/3$ & $0$ & $0$ \\ 
  $c_{BB}$ &  $(y+\frac{1}{3})^2$
          &  $\frac{1}{3} (y+\frac{1}{3})^2$
          &  $\frac{1}{2} (y+\frac{1}{3})^2$ 
          &  $\frac{1}{6} (y+\frac{1}{3})^2$ \\ 
    $c_{GG}$ & $1$ & $0$ & $1/2$ & $0$ \\
    $c_{\Phi2}$ & $1$ & $1/3$ & $1/2$ & $1/6$ \\
    $c_{\Phi3}$ & $1$ & $1/3$ & $1/2$ & $1/6$ \\
    $c_{bb}$ & $2$ & $2$ & $1$ & $1$ \\
    $c_{bB}$ & $2(y- \frac{1}{6})$ & $2 (y-\frac{1}{6})$ &
               $y-\frac{1}{6}$ &  $y-\frac{1}{6}$ \\
    $c_{bG}$ & $2$ & $6$ & $1$ & $3$ \\[0.2cm] \hline
\end{tabular}
\end{center} \par
\noindent
\begin{center}
\begin{tabular}{|c|c|c|c|} \hline
\multicolumn{4}{|c|}{Table 5: Models of Set II.} 
\\ \hline
\multicolumn{1}{|c|}{Model} &
 \multicolumn{1}{|c|}{New Particle} &
  \multicolumn{1}{|c|}{$I$} &
    \multicolumn{1}{|c|}{$Y$} \\ \hline
  4 & Scalar $\Psi$ & $1/2$ & $1/2$   \\  
  5 &  $F_L,~F_R$ & $1/2$ & $1/6$ \\
  6A &  $F_L,~F_R$ & $0$ & $-1/3$ \\
  6B &  $F_L,~F_R$ & $0$ & $2/3$ \\ \hline
\end{tabular}	
\end{center}
\noindent


\end{document}